\def\CHfour{{\fam0 CH_4}}
\def\CHthree{{\fam0 CH_3}}

\def\HtwoCN{{\fam0 H_2CN}}

\def\HCN{{\fam0 HCN}}

\def\CHthreeOH{{\fam0 CH_3OH}}
\def\CHthreeO{{\fam0 CH_3O}}

\def\CHtwoOH{{\fam0 CH_2OH}}

\def\COtwo{{\fam0 CO_2}}

\def\CthreeHtwo{{\fam0 C_3H_2}}
\def\CthreeHthree{{\fam0 C_3H_3}}

\def\CtwoHtwo{{\fam0 C_2H_2}}

\def\CtwoHfour{{\fam0 C_2H_4}}

\def\CtwoHsix{{\fam0 C_2H_6}}

\def\oneCHtwo{{\fam0 ^{1}CH_2}}

\def\hd189{{\fam0 HD 189733b}}
\def\hd209{{\fam0 HD 209458b}}

\def\HtwoCO{{\fam0 H_2CO}}

\def\HtwoO{{\fam0 H_2O}}
\def\Htwo{{\fam0 H_2}}
\def\Hatom{{\fam0 H}}

\def\M{{\fam0 M}}
\def\NO{{\fam0 NO}}
\def\NH{{\fam0 NH}}
\def\NHtwo{{\fam0 NH_2}}
\def\NHthree{{\fam0 NH_3}}
\def\Net{{\fam0 Net:}}
\def\N{{\fam0 N}}
\def\Ntwo{{\fam0 N_2}}

\def\NtwoHtwo{{\fam0 N_2H_2}}
\def\HtwoNN{{\fam0 H_2NN}}
\def\NtwoHthree{{\fam0 N_2H_3}}
\def\NtwoHfour{{\fam0 N_2H_4}}
\def\OH{{\fam0 OH}}

\def\Oatom{{\fam0 O}}

\def\cmtwo{{\fam0 cm^2}}

\def\PHthree{{\fam0 PH_3}}

\def\scinot#1.{\hbox{$\,$ $\times$ $10^{#1}$}}
\def\smone{{\fam0\,s^{-1}}}

\def\ten#1.{\hbox{$10^{#1}$}}

\def\deg{\ifmmode^\circ\else$\null^\circ$\fi}
\def\spose#1{\hbox to 0pt{#1\hss}}
\def\lta{\mathrel{\spose{\lower 3pt\hbox{$\mathchar "218$}}\raise 2.0pt\hbox{$\mathchar"13C$}}}
\def\gta{\mathrel{\spose{\lower 3pt\hbox{$\mathchar "218$}}\raise 2.0pt\hbox{$\mathchar"13E$}}}
\def\lrarrow{\mathrel{\spose{\lower 1pt\hbox{$\rightarrow$}}\raise 3.0pt\hbox{$\
leftarrow$}}}
\newcount\citeno
\def\cnum{
  \global\advance\citeno by 1
  \the\citeno}
\documentclass{rsauthor}
\usepackage{times}
\usepackage{mathptm}

\begin{document}


\title{Chemical Kinetics on Extrasolar Planets}

\author{Julianne I. Moses}
\address{Space Science Institute, 4750 Walnut Street, Suite 205, Boulder, CO, 80301, USA}
\corres{jmoses@spacescience.org}

\date{}

\abstract{Chemical kinetics plays an important role in controlling the atmospheric composition 
of all planetary atmospheres, including those of extrasolar planets.  For the hottest 
exoplanets, the composition can closely follow thermochemical-equilibrium predictions, at least 
in the visible and infrared photosphere at dayside (eclipse) conditions.  However, for atmospheric 
temperatures $\lta$ 2000 K, and in the uppermost atmosphere at any temperature, chemical kinetics 
matters.  The two key mechanisms by which kinetic processes drive an exoplanet atmosphere out 
of equilibrium are photochemistry and transport-induced quenching.  We review these disequilibrium 
processes in detail, discuss observational consequences, and examine some of the current evidence 
for kinetic processes on extrasolar planets.}



\jname{Phil. Trans. Roy. Soc.~A}
\maketitle

\section{Introduction}

In two short decades, the study of extrasolar planets has evolved from a relatively 
esoteric, theory-driven, scientific diversion to a full-fledged, observation-driven, vibrant 
profession 
\cite{deming09,seagerbook}.
With 859 exoplanets discovered as of January 2013 
\cite{encyclo}
and thousands more {\it Kepler\/} planetary candidates waiting to be confirmed 
\cite{borucki11,batalha12},
our Sun has lost its status as a unique planetary host.  However, 
each planetary system discovered to date displays its own unique properties, and the sheer 
diversity of those properties is staggering.  Most of the known exoplanets are expected to 
possess atmospheres of some sort, but those atmospheres will be equally diverse and often have 
no Solar-System analogs.  Can we use basic physical and chemical principles to predict the 
characteristics and behavior of exoplanet atmospheres?  How good are our predictions, and what 
do we learn from comparisons of observations and theory?

Current observational techniques used to analyze exoplanet atmospheres include direct imaging 
[6-8] 
and transit and eclipse measurements 
[9-12].  
Because exoplanets are so much fainter than their host stars, atmospheric 
characterization by direct imaging is so far only viable for young, bright, hot planets that orbit
relatively far from their host stars, and concrete information on atmospheric composition has only 
recently been reported from this technique
[13-21]. 
Transit and eclipse observations, where light from the 
system is observed to dim as the planet passes in front of and behind the host star as seen from 
the observer, have been more fruitful in determining atmospheric composition, but observational 
biases still limit the types of planets than can be studied in this manner.  Transits have a higher 
probability of being observed for planets orbiting very close to their host stars, and the signal is 
stronger for larger planets.  Therefore, close-in giant planets --- the so-called ``hot Jupiters'' 
and ``hot Neptunes'' --- currently dominate exoplanet atmospheric observations, and our review of 
exoplanet chemistry will focus on such planets.  Transit and eclipse observations have enabled the
first-ever detections of neutral and ionized atoms like Na, H, O, C$^+$, Si$^{++}$, Mg$^+$, and K 
[9,22-26] 
and molecules like $\HtwoO$, $\CHfour$, $\COtwo$, and CO 
[27-36] 
in exoplanet atmospheres.

The first-order properties of exoplanet atmospheres can be predicted theoretically based on the 
planet's mass, internal heat flux (and/or age), assumed bulk elemental composition, and incident 
stellar flux \cite{marley07}.  
Given these basic parameters, models of thermochemical equilibrium, 
cloud condensation, and radiative and convective energy transport can be used to predict the 
composition and thermal structure within the planet's atmosphere, as well as with the observable 
spectral behavior
[38-46]. 
Chemical equilibrium is a convenient starting point for these types of calculations, as the 
abundances of individual species can be calculated in a straightforward manner for any given 
temperature, pressure, and elemental composition through minimization of the Gibbs free energy of the 
system 
\cite{burrows99,lodders02}.  
The actual pathways (e.g., chemical reactions) required to achieve that equilibrium do 
not matter, nor does the history of the system, greatly simplifying the calculations.  At high 
temperatures, such as within the deep atmospheres of hot Jupiters, chemical reactions can overcome 
energy barriers to proceed equally well in both the forward and reverse direction, and the 
chemical equilibrium assumption is justified.  However, for cooler regions of the atmosphere or 
when incident ultraviolet photons or high-energy ionizing/dissociating particles are present, 
chemical equilibrium becomes more difficult to achieve, and chemical kinetics or other disequilibrium
mechanisms can control the composition.  

The two main kinetics-related disequilibrium processes expected to modify abundances within exoplanet 
atmospheres are {\em photochemistry\/} and {\em transport-induced quenching}.  Photochemistry refers 
to the chemical kinetics that results from the absorption of short-wavelength stellar photons or 
high-energy corpuscular radiation like cosmic rays.  Transport-induced quenching refers to the 
mechanism by which the 
atmospheric composition is driven away from chemical equilibrium as a result of the dominance of 
transport processes like convection or large-scale ``eddy'' diffusion (e.g., from gravity-wave breaking 
or wave-driven circulation) over chemical reactions  
[47-49].  
While often 
considered second-order effects, both of these disequilibrium processes can significantly alter the 
composition and hence the radiative properties, thermal structure, and even dynamics of the atmosphere, 
making them important processes to consider in theoretical models.  As an example, the photolysis of 
methane on Jupiter and the other giant planets within our own Solar System leads to the generation of 
complex hydrocarbons, introducing numerous trace photochemical products to the Jovian stratosphere 
[e.g., 50].  
Some of these products, like $\CtwoHtwo$ and $\CtwoHsix$, become key infrared 
coolants 
\cite{yelle01}, 
while some of the more refractory products can condense in the 
stratosphere to form hazes that absorb sunlight and provide localized heating 
\cite{kark05}.  
The photochemical 
products strongly influence the spectral behavior, radiative energy transport, and thermal profile, 
which feed back to affect the stratospheric circulation 
[e.g., 53].  
Due to the intense 
ultraviolet flux incident onto the atmospheres of close-in transiting exoplanets, photochemistry is 
expected to play a significant role on hot Jupiters 
[54-66],  
 although the observable consequences may be 
restricted to high altitudes in some cases, particularly for the hotter atmospheres. 
Transport-induced quenching can potentially affect the composition throughout the visible and infrared
photosphere of exoplanets 
[42,62-63,66-72],
thereby affecting the thermal structure and spectral behavior.

We review current knowledge of these two disequilibrium processes and discuss how atmospheric properties 
like the thermal structure, bulk elemental abundance, and transport properties can affect the predicted 
composition.  We also briefly discuss the observational evidence for disequilibrium compositions. 

\section{Transport-induced quenching}

Transport-induced quenching within giant-planet atmospheres was first described by Prinn and Barshay 
\cite{prinn77} 
to explain the observed $\sim$1 part per billion (ppb) abundance of CO in Jupiter's upper troposphere, 
despite the negligible amount expected at local temperature-pressure conditions from 
chemical-equilibrium arguments. As a parcel of gas is transported through the atmosphere, the mole 
fraction of a constituent can become ``quenched'' when transport time scales drop below the chemical 
kinetics time scales required to maintain that constituent in equilibrium with other species.  The 
mole fraction of that species then remains fixed at that quenched abundance.  Quenching 
typically occurs when the temperature of the system drops low enough that reactions no longer occur 
equally well in both the forward and reverse directions; however, pressure changes can also play a role.  
The best-studied quenching process is that of CO-$\CHfour$ interconversion due to vertical
mixing because of the importance of that process
for Solar-System giant planets 
[47,73-77],
brown dwarfs 
[70-71,78-84],
and extrasolar giant planets
[17,60,62-63,67,69,85-86]. 
In fact, CO-$\CHfour$ quenching has particularly important observational consequences for hot Jupiters 
and brown dwarfs because their atmospheric temperature profiles often cross the boundary between regions 
in which $\CHfour$ or CO are the dominant carbon constituents, with transport-induced 
quenching then leading to CO and/or CH$_4$ mole fractions that are orders of magnitude different from 
chemical-equilibrium predictions.  The resulting spectral implications can be major (see above references).  
Although time-constant arguments have historically been used to predict the quenched abundances of CO and 
$\CHfour$, B\'ezard et al.~\cite{bezard02}, 
Moses et al.~\cite{moses11}, 
and Visscher and Moses \cite{visscher11} 
discuss some of the pitfalls of such techniques that can arise if the underlying assumptions 
are not sound. Many investigators instead now solve the full continuity equations for a large number 
of species to track transport-induced quenching 
[61-63,65-66,71-72,77]
a technique that has the added benefit of more accurately predicting the profiles of other potentially 
spectrally active species (e.g., $\COtwo$, $\CtwoHtwo$, HCN) once some critical ``parent'' molecules are quenched.

However, these more sophisticated models are also only as good as their inputs and assumptions, 
and although the kinetics of the C-H-O system has been well studied due to combustion-chemistry 
and terrestrial atmospheric-chemistry applications 
[87-89], 
the exact mechanism involved with $\CHfour$-CO quenching in reducing environments is not strictly 
established.  As originally hypothesized \cite{prinn77}, 
individual reactions that convert oxidized carbon (e.g., CO, $\COtwo$, HCO, $\HtwoCO$, $\CHtwoOH$, $\CHthreeO$, 
$\CHthreeOH$) to reduced carbon (e.g., $\CHfour$, $\CHthree$, $\CtwoHtwo$, $\CtwoHfour$, $\CtwoHsix$) and 
{\em vice versa\/} are of critical importance to the problem, as reactions within the oxidized or reduced 
families tend to be faster.  The reaction $\Htwo$ + $\HtwoCO$ $\rightleftharpoons$ OH + $\CHthree$ was 
originally proposed as the rate-limiting step for CO-$\CHfour$ conversion 
\cite{prinn77,lewis84,fegley94}, 
but this reaction is likely too slow under typical brown-dwarf or giant-planet conditions to be a key 
player in the dominant interconversion mechanism 
[62,71,74,76-77,80].  
Yung et al.~\cite{yung88} 
suggest that the reaction H + $\HtwoCO$ + M $\rightarrow$ $\CHthreeO$ + M is the 
rate-limiting step as a potential bottleneck to the necessary conversion of the strong CO bond to a weaker 
single-bonded C-O species.  However, more recent studies point to alternative reactions being the rate-limiting 
steps; the most thorough and updated discussions of the problem are presented in 
\cite{moses11,visscher11}. 

\begin{figure}
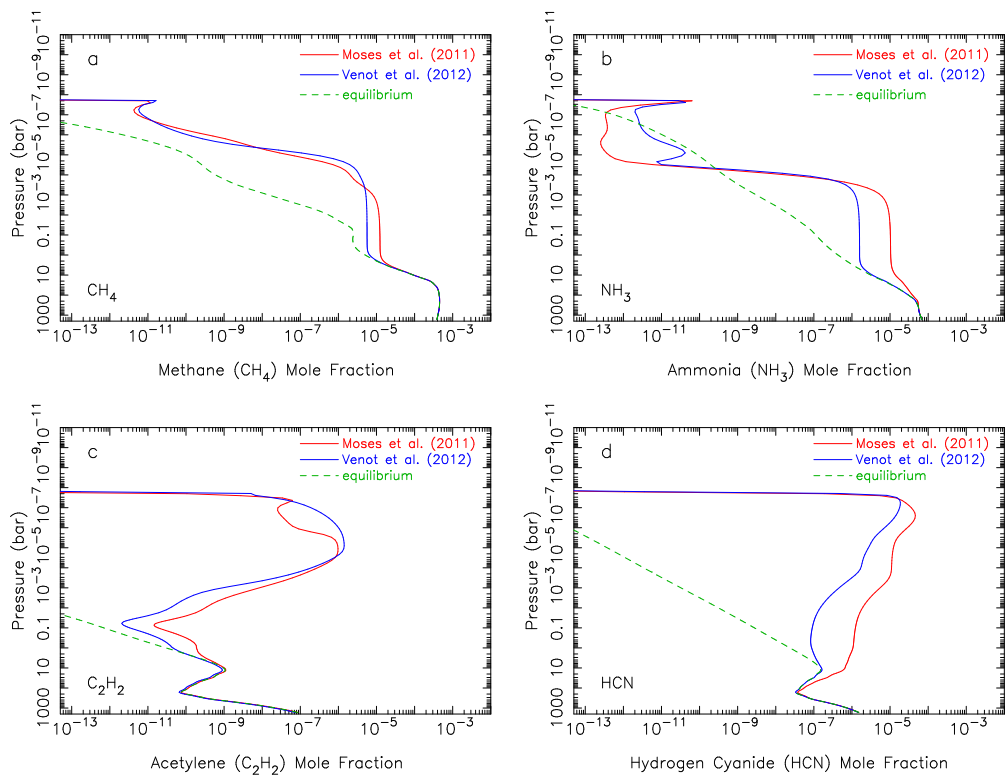

\begin{tabular}{ll}
{\includegraphics[angle=-90,clip=t,scale=0.28]{fig1_topleft.ps}} 
& 
{\includegraphics[angle=-90,clip=t,scale=0.28]{fig1_topright.ps}} 
\\
{\includegraphics[angle=-90,clip=t,scale=0.28]{fig1_bottomleft.ps}} 
& 
{\includegraphics[angle=-90,clip=t,scale=0.28]{fig1_bottomright.ps}} 
\\
\end{tabular}
\caption{Mole fractions (volume mixing ratios) of (a) CH$_4$, (b) NH$_3$, (c) C$_2$H$_2$, and (d) HCN 
from chemical models of HD 189733b, assuming thermochemical equilibrium (green dashed curves), or 
including photochemical kinetics and transport with the Moses et al.~\cite{moses11} 
reaction mechanism (red solid curves) or with the Venot et al.~\cite{venot12} reaction mechanism 
(blue solid curves).  The dayside thermal structure and nominal eddy diffusion coefficient profile from 
\cite{moses11} 
were used  throughout the modeling 
[see also 66,90].
Note that the kinetics models begin to diverge from the equilibrium profiles at different depths 
due to differences in the adopted reaction mechanism.  The resulting ``quenched'' abundances therefore 
differ between models, despite the same assumptions concerning atmospheric transport.  See Venot et 
al.~\cite{venot12} for similar figures.  Online version is in color.\label{quenchcomp}}
\end{figure}

Figure \ref{quenchcomp} demonstrates that the quenched abundances of species like $\CHfour$ and $\NHthree$ 
are very sensitive to the individual reactions and rate coefficients in the adopted reaction mechanism 
for these kinetics-transport models.  Moreover, once molecules like $\CHfour$ and $\NHthree$ quench, the 
abundances of other species such as $\CtwoHtwo$ and HCN can be affected, and the profiles of minor species 
can become complicated.  Because the full reaction mechanisms for the Moses et al.~\cite{moses11} 
and Venot et al.~\cite{venot12} 
studies are published online, intercomparisons between these models is relatively 
straightforward; therefore, we focus our discussion on these two models and apply the mechanisms to the 
dayside atmosphere of HD 189733b.  Venot et al.~\cite{venot12} 
do not provide a detailed discussion of their dominant $\CHfour$ $\rightarrow$ CO quench mechanism, but it is 
clear that interconversion between these two species is more efficient in their model than in Moses et 
al.~\cite{moses11}, as the CH$_4$ mole fraction in the Venot et al.~model continues to follow the equilibrium 
predictions to higher altitudes (lower pressures/temperatures) than in \cite{moses11}. 
The resulting quenched $\CHfour$ abundance is therefore lower in the Venot et al.~\cite{venot12} 
model than in the Moses et al.~\cite{moses11} model.  

A detailed examination of the two mechanisms reveals that the main difference 
with respect to the $\CHfour$ quench behavior derives from the adopted rate coefficients for the 
reaction H + $\CHthreeOH$ $\rightarrow$ $\CHthree$ + $\HtwoO$.  The Venot et al.~mechanism originates
from combustion-chemistry studies that use the Hidaka et al.~\cite{hidaka89} 
rate-coefficient estimate for this reaction; some previous giant-planet quenching studies also 
initially adopted this rate coefficient \cite{visscher10co}.  
However, as is discussed by Visscher et al.~\cite{visscher10co} and Moses et al.~\cite{moses11}, 
the H + $\CHthreeOH$ $\rightarrow$ $\CHthree$ + $\HtwoO$ reaction 
likely proceeds with a very large energy barrier and is therefore much slower than was estimated by 
\cite{hidaka89}.  
When a more realistic rate coefficient for this reaction, as calculated from {\em ab initio\/} 
transition-state theory \cite{moses11}, is incorporated into the Venot et al.~mechanism, 
the resulting quenched CH$_4$ mole fraction is more in line with that of Moses et al.~\cite{moses11}, 
with remaining differences being largely due to the adoption of a larger rate coefficient for 
$\HtwoO$ + $\oneCHtwo$ $\rightarrow$ $\CHthreeOH$ in the Venot et al.~model (which was again seems to 
be based on an 
estimate [e.g., 91]\,) 
than in the Moses et al.~model (where the rate-coefficient calculations of \cite{jasper07} have been adopted).  

Although we think that CH$_4$-CO quenching is unlikely to proceed exactly as is described by the Venot et 
al.~mechanism, due to some problems with individual rate coefficients discussed above, it is possible 
that other reactions and/or rate coefficients not considered by either \cite{moses11} or \cite{venot12} 
(or any other investigations) could be dominating on hot Jupiters.  
As an example, Hidaka et al.~\cite{hidaka89} favored a fast rate for H + $\CHthreeOH$ $\rightarrow$ 
$\CHthree$ + $\HtwoO$ precisely because they were looking for an effective way to reproduce the observed 
yield of $\CHfour$ in their methanol decomposition experiments.  If that reaction is not responsible, 
as we indeed suggest it is not, then other reactions must be taking up the slack, and it remains to be 
demonstrated whether any of the updated reaction rate coefficients in the Moses et al.~\cite{moses11}, 
Venot et al.~\cite{venot12} mechanisms, or those of other investigations \cite{zahnle09sulf,line11gj436b,koppa12}, 
can reproduce all available experimental data.  The quenched abundance of $\CHfour$ should therefore 
be considered only accurate to within a factor of $\sim$2 due to kinetics uncertainties, with other factors 
such as atmospheric transport properties and thermal structure contributing additional uncertainty.  It is 
obvious from Fig.~\ref{quenchcomp}, however, that CO-$\CHfour$ quenching matters on HD 189733b and likely 
other close-in hot Jupiters, with potential observable consequences, just as on brown dwarfs and 
directly-imaged extrasolar giant planets 
[13,17,70,79,81]. 
The colder the exoplanet atmosphere, the deeper the quench point, and the larger the quenched $\CHfour$ mole 
fraction that could be present in the visible-infrared photosphere \cite{moses11}.

Similarly, interconversion between the main nitrogen species $\Ntwo$ and $\NHthree$ can be kinetically 
inhibited in planetary and brown-dwarf atmospheres 
[49,94-97],
leading to quenching of both species.  The $\Ntwo$-$\NHthree$ quench point likely occurs deeper in the 
atmosphere than the CO-$\CHfour$ quench point, but the kinetics of nitrogen species under high-temperature, 
high-pressure, reducing conditions is even more uncertain than that of carbon and oxygen.  The key to the 
kinetics of interconversion of $\Ntwo$ and $\NHthree$ is likely the reduction of the strongly triple-bonded 
$\Ntwo$ into progressively weaker N-N bonded N$_2$H$_x$ species, followed by thermal decomposition or 
disproportionation reactions converting the N$_2$H$_x$ molecule into two NH$_x$ species.  Initial 
suggestions for the rate-limiting steps in the process 
[75,94-95],
such as $\Ntwo$ + $\Htwo$ $\rightleftharpoons$ $2\, $NH have given way to other suggestions because 
the above reaction will 
likely be too slow to play any significant role in the $\Ntwo$-$\NHthree$ conversion \cite{moses10}.  
Alternative suggestions for the rate-limiting step in NH$_3$ quenching include (a) $\NtwoHthree$ + M 
$\rightleftharpoons$ H + $\NtwoHtwo$ + M \cite{moses10}, where M is any additional atmospheric species in 
this three-body reaction, (b) $2\, \NHtwo$ $\rightleftharpoons$ $\NtwoHtwo$ + $\Htwo$ \cite{line11gj436b}, 
and (c) NH + $\NHtwo$ $\rightleftharpoons$ $\NtwoHtwo$ + H \cite{moses11}.  These are all viable possibilities, 
and published rate coefficients for these reactions or their reverses are available from theoretical 
calculations 
[98-101] 
or experiments \cite{stoth95}, 
but large uncertainties
and discrepancies for these and other potentially important reaction rate coefficients exist in the literature.  
The predicted quenched abundance of $\NHthree$ on HD 189733b differs by about an order of magnitude when 
using the Venot et al.~\cite{venot12} 
versus the Moses et al.~\cite{moses11} 
reaction mechanism (see Fig.~\ref{quenchcomp} and \cite{venot12}), 
and Venot et al.~\cite{venot12} have tested several other 
combustion-based reaction mechanisms, all which predict different quenched $\NHthree$ abundances on 
HD 189733b.  These differences emphasize the kinetic uncertainties that currently plague quench 
predictions for nitrogen species.

Kinetic interconversion between $\Ntwo$ and $\NHthree$ is clearly much more effective in models that use
the Venot et al.~\cite{venot12} mechanism as opposed to the Moses et al.~\cite{moses11} 
mechanism.  Ammonia diverges from the equilibrium profile much deeper with the Moses et al.~\cite{moses11} 
mechanism, and although 
the $\NHthree$ does not truly quench until higher altitudes due to transport time scales being only slightly 
smaller than the kinetic conversion time scales in the nearly isothermal region between a few bars and 
$\sim$0.1 kbar in the HD 189733b model \cite{moses11}, 
the resulting quenched NH$_3$ abundance is significantly greater in the Moses et al.~\cite{moses11} 
model than the Venot et al.~\cite{venot12} 
model.  The main differences appear to be related to reactions of $\NHtwo$ with $\NHtwo$ and/or 
$\NHthree$.  Venot et al.~\cite{venot12} have adopted very large rate coefficients for some reactions 
that convert $\NHtwo$ to N$_2$H$_x$ species based 
on the work of Konnov and De Ruyck \cite{konnov00,konnov01}.  For example, the Venot et al.~ rate 
coefficient for $\NHtwo$ + $\NHthree$ $\rightarrow$ $\NtwoHthree$ + $\Htwo$ (based on \cite{konnov00}) 
is more than three orders of magnitude larger at 1600 K than the expression used in Moses et 
al.~\cite{moses11} (based on \cite{dean84}), 
and this reaction plays a major role in the efficient interconversion of $\NHthree$ 
$\rightleftharpoons$ $\Ntwo$ in the Venot et al.~\cite{venot12} 
mechanism.  However, the rate-coefficient expression advocated by Konnov \& De Ruyck \cite{konnov00} 
is simply an {\it ad hoc\/} eight-fold reduction of a previous estimate \cite{dove79}, 
which Konnov \& De Ruyck altered to keep that reaction from having 
any adverse effects on their experimental simulations \cite{konnov00}.  
As such, the Konnov \& De Ruyck expression is more of an upper limit than a true recommendation.  
In fact, the $\NHtwo$ + $\NHthree$ $\rightarrow$ $\NtwoHthree$ + $\Htwo$ reaction likely has a 
significantly higher barrier than the Konnov \& De Ruyck expression indicates, and the reaction 
is not expected to be important under conditions relevant to hydrazine combustion or ammonia pyrolysis 
\cite{dean84}.  It is also unlikely to be important for NH$_3$-$\Ntwo$ quenching in giant-planet 
atmospheres if the rate coefficient has an energy barrier similar to that suggested by Dean et 
al.~\cite{dean84} (see also \cite{moses11}); however, 
further rate-coefficient information on this reaction is needed before it can truly be ruled out as a 
participant in the $\NHthree$ $\rightleftharpoons$ $\Ntwo$ interconversion process.  The adoption of a fast 
rate coefficient for this reaction strongly influences the low derived $\NHthree$ abundance in the Venot 
et al.~model.

The $\NHtwo$ + $\NHtwo$ $\rightarrow$ $\NtwoHtwo$ + $\Htwo$ reaction represents a more interesting case, 
as the product pathways and rate coefficients are not well known for the $\NHtwo$ + $\NHtwo$ reaction.  
Venot et al.~\cite{venot12} 
adopt a relatively large rate coefficient for this reaction based on 
Konnov \& De Ruyck \cite{konnov01}, 
which in turn was influenced by the experimental study of Stothard et al.~\cite{stoth95} that 
indicates that $\NtwoHtwo$ (as an unidentified isomer) is an important product 
in the $\NHtwo$ + $\NHtwo$ reaction at room temperature.  Theoretical models 
[98-99,101]
in contrast, suggest that the standard pathways producing $\Htwo$ + $\NtwoHtwo$ (various isomers) 
are relatively unimportant at room temperature.  In the Klippenstein et al.~\cite{klipp09} calculations, for
instance, two NH$_2$ radicals interact on a singlet potential energy surface via barrierless addition (i.e., 
to form $\NtwoHfour$), with the stabilization of the adduct dominating especially at low temperatures and 
high pressures, or potential elimination of H and $\Htwo$ occurring at higher temperatures (e.g., to form 
H + $\NtwoHthree$ or $\Htwo$ + various isomers of $\NtwoHtwo$).  On the triplet surface, the $\NHtwo$ + 
$\NHtwo$ reaction can occur via hydrogen abstraction to form NH + $\NHthree$ \cite{klipp09}, which is 
especially important at high temperatures.  The barriers 
leading to $\Htwo$ + $\NtwoHtwo$ are sufficiently high on the singlet surface in the Klippenstein et 
al.~\cite{klipp09} models that the product channels forming various isomers of $\NtwoHtwo$ are unimportant 
even at the moderately high temperatures relevant to $\NHthree$ quenching on the giant planets (1500-2100 K).  
Moses et al.~\cite{moses11} have adopted the rate-coefficient recommendations of Dean \& 
Bozzelli~\cite{dean00} for the $\Htwo$ elimination pathways, which are more in line with the Klippenstein 
et al.~\cite{klipp09} recommendations than the Konnov \& De Ruyck \cite{konnov01} recommendations, and 
the resulting $\NHtwo$ + $\NHtwo$ $\rightarrow$ $\NtwoHtwo$ + $\Htwo$ or $\rightarrow$ $\HtwoNN$ + $\Htwo$ 
pathways are not very important in $\NHthree$ $\rightarrow$ $\Ntwo$ conversion in their HD 189733b and HD
209458b models.  On the other hand, Asatryan et al.~\cite{asatryan10} suggest that there could be a
potential low-energy pathway to $\NHtwo$ + $\NHtwo$ $\rightarrow$ $cis$-$\NtwoHtwo$ + $\Htwo$ that could 
occur through an ``activated'' $\NtwoHfour$ intermediate even at moderately low temperatures.
Stereoselective attack of $cis$-$\NtwoHtwo$ by $\Htwo$ then leads to $\Ntwo$ + 2$\Htwo$ with a comparatively 
low barrier.  As mentioned by Altinay \& Macdonald~\cite{altinay12}, this suggested pathway needs further 
theoretical verification, but if plausible, it  might lead to a more efficient means to $\NHthree$ 
$\rightarrow$ $\Ntwo$ conversion in exoplanet atmospheres than has been implemented in the Moses et al.~model.  

In all, the Moses et al.~\cite{moses11} and Venot et al.~\cite{venot12} mechanisms can be considered to 
bracket the low and high ends of possible $\NHthree$ $\leftrightharpoons$ $\Ntwo$ conversion efficiencies, and 
the expected quenched $\NHthree$ abundance on transiting exoplanets is uncertain by about an order 
of magnitude (see also Venot et al.~\cite{venot12}).  To improve that situation, we need reliable rate 
coefficients for several reactions 
involving NH$_x$ and N$_2$H$_x$ species at temperatures of $\sim$1500-2200 K and pressures of 
$\sim$1-1000 bar.  Of particular interest are rate coefficients for the various possible pathways 
involved with the reactions of NH$_2$ + NH$_2$, NH$_2$ + NH$_3$, NH + NH$_2$, NH + NH$_3$, $\Htwo$ + 
$\NtwoHtwo$ (various isomers), $\Htwo$ + $\NtwoHthree$, H + N$_2$H$_x$, and thermal decomposition of 
$\NtwoHfour$, $\NtwoHthree$, and various $\NtwoHtwo$ isomers.  Since the chemistry of ammonia and 
hydrazine is also important on Jupiter and Saturn, extension to $\sim$100 K temperatures would also 
be valuable for investigation of tropospheric photochemistry on those planets.

Carbon monoxide and molecular nitrogen also technically quench when CO-$\CHfour$ and $\Ntwo$-$\NHthree$ 
interconversion reactions cease to be able to compete with vertical transport processes. However, because
these molecules are the dominant carbon and nitrogen constituents at the quench points in the HD 189733b and 
HD 209458b models \cite{moses11,venot12}, the quenching behavior for these species is less obvious since 
their equilibrium mole fractions are expected to remain constant to higher altitudes anyway.  If the quench 
points on exoplanets were to occur in a region where both the $\CHfour$ and CO (or $\Ntwo$ and NH$_3$) have 
similar abundances, then the quenching of both species would be more obvious.

Once major molecules like NH$_3$ and CH$_4$ are quenched, the abundance of other species can be 
strongly affected.  For example, Fig.~\ref{quenchcomp} illustrates that the HCN and $\CtwoHtwo$ 
abundances depart from equilibrium when $\NHthree$ and $\CHfour$ quench on HD 189733b, and 
these molecules continue to react with the quenched ``parent'' molecules up to higher altitudes, 
causing their column abundances to greatly exceed equilibrium predictions.  Acetylene maintains a 
pseudo-equilibrium with $\Htwo$ and quenched $\CHfour$ up $\sim$0.1 bar in this model (see 
\cite{moses11} for more information about the important reaction schemes).  The large bulge in the 
$\CtwoHtwo$ abundance at higher altitudes is due to photochemical and disequilibrium thermochemical 
kinetics processes.  Similarly, HCN 
maintains a pseudo-equilibrium with quenched NH$_3$ and CH$_4$ (with a net reaction 
$\NHthree$ + $\CHfour$ $\rightleftharpoons$ HCN + 3$\Htwo$, see Moses et al.~\cite{moses11} for 
detailed reaction schemes) to $\sim$0.1 mbar, above which the downward transport of photochemically 
produced HCN dominates the profile.  Transport-induced quenching can therefore affect species 
abundances even before their own quench points are reached, and full kinetics models are needed 
to describe this behavior.  

Quenching does not occur solely in the vertical direction.  Possible horizontal quenching was first 
discussed by Cooper \& Showman~\cite{coop06}.  Time-constant arguments \cite{coop06,moses11} 
indicate that horizontal winds have the potential to transport gas species to cooler atmospheric 
regions before the species have time to equilibrate, which could particularly affect transit 
observations of the cooler terminator regions. Similarly, quenching (vertical or horizontal) can 
affect the atmospheric composition of close-in eccentric exoplanets \cite{visscher12} such that 
the large swings in temperature over the orbit may not necessarily be accompanied by large changes 
in atmospheric composition if orbital time scales are shorter than chemical reaction time scales.  
Ag{\'u}ndez et al.~\cite{agundez12} have followed up these ideas with a more sophisticated kinetics model 
to investigate horizontal quenching on HD 209458b under the simplifying assumptions of constant 
zonal winds as a function of altitude, latitude, and longitude, with no photochemistry or vertical 
transport included; they find that horizontal quenching is indeed very important, especially at 
pressures less than $\sim$0.1 bar (depending on the species).  The horizontal quenching suppresses 
longitudinal abundance gradients, greatly reducing dayside vs. terminator abundance differences, 
with resulting major observational consequences.  Essentially, Ag{\'u}ndez et al.~\cite{agundez12} 
find that abundances above the $\sim$0.1 bar level are quenched horizontally to dayside equilibrium 
values, whereas quenching due to vertical mixing controls the abundances at deeper levels down to 
the quench point.  The resulting terminator and nightside abundance profiles differ from 
models based on either thermochemical equilibrium or kinetics considering vertical transport only, 
and the main prediction is a reduction in the mid-to-high altitude abundance of species like CH$_4$ 
and NH$_3$ that are relatively unstable at the high dayside temperatures of HD 209458b but that 
would nominally be stable at cooler terminator and nightside temperatures.  

Clearly, the chemical profiles in the 3D situation will be complicated, and kinetics and transport-induced 
quenching in both the vertical and horizontal direction must be considered to accurately reflect the 
situation in the real atmosphere.  The inclusion of vertical 
mixing in such longitudinally variable models is an important next step, as it would more realistically 
capture the dayside abundances of vertically quenched species like NH$_3$ and CH$_4$, and thus more 
accurately predict the horizontally quenched terminator and nightside abundances.  The addition of 
photochemistry to the problem will complicate things further, as photochemical time constants tend to 
be very short in the upper atmospheres of these planets.  Ultimately, it will be valuable to move 
toward the inclusion of more sophisticated kinetics directly within the 3D general circulation models, 
as the disequilibrium constituents can affect the radiative properties of the atmosphere, and the 
resulting feedback on the temperatures and even dynamics could be important.

\begin{figure}
\centering
\includegraphics[angle=-90,scale=0.37]{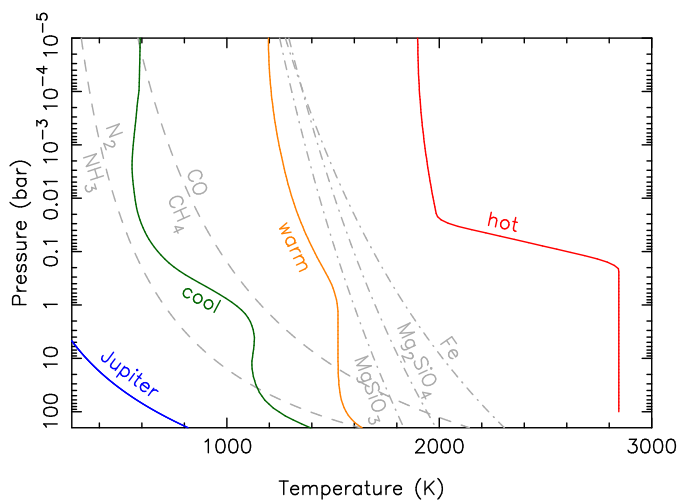}
\caption{Thermal profiles for the hypothetical ``hot'', ``warm'', and ``cool'' exoplanets (as labeled) 
used in the chemical models shown in Fig.~\ref{figmix}.  The gray dashed lines represent the 
equal-abundance curves for CH$_4$-CO and NH$_3$-N$_2$.  Profiles to the right of these curves are 
within the N$_2$ and/or CO stability fields.  The dot-dashed lines show the condensation curves for 
MgSiO$_3$, Mg$_2$SiO$_4$, and Fe (solid, liquid) \cite{visscher10rock}.  Online version is in
color.\label{figtemp}}
\end{figure}

\section{Photochemistry}

The consequences of photochemistry on the atmospheric composition of exoplanets have been explored by 
numerous groups 
[54-66,109].  
We briefly review how photochemistry affects the carbon, nitrogen, and 
oxygen species in the 0.1-1000 mbar region of hot Jupiters and hot Neptunes, i.e., the pressure region 
at which the infrared transit and eclipse observations have their highest sensitivity.  Higher-altitude 
thermospheric photochemistry is discussed by 
[56-58], 
and sulfur photochemistry 
is discussed by \cite{zahnle09sulf}.  Little work has yet been done on the photochemistry of other elements 
like phosphorus, alkalis, or metals due to a lack of relevant kinetic information.

Although stellar UV photons are the ultimate instigators of photochemistry, the resulting 
abundances of the observable photochemical products on hot Jupiters depend less on the magnitude of 
the incident UV flux than on the overall atmospheric thermal structure (and hence the stellar flux in 
the visible and near-IR) because of the high sensitivity of the ``parent'' molecule abundances to 
temperature.  Near-solar composition atmospheres that are hotter than $\sim$1600 K in the 0.1-1000 mbar 
region will contain CO, H$_2$O, and $\Ntwo$ as their dominant heavy molecular constituents.  These 
molecules are relatively stable against photochemical destruction and/or tend to recycle efficiently 
(e.g., \cite{moses11}) so that photochemistry does not remove them from the 0.1-1000 mbar region (unless 
temperatures are so high that these molecules, too, become unstable).  Cooler atmospheres will contain 
more CH$_4$ and NH$_3$, which are more interesting from a photochemical standpoint.

\begin{figure}
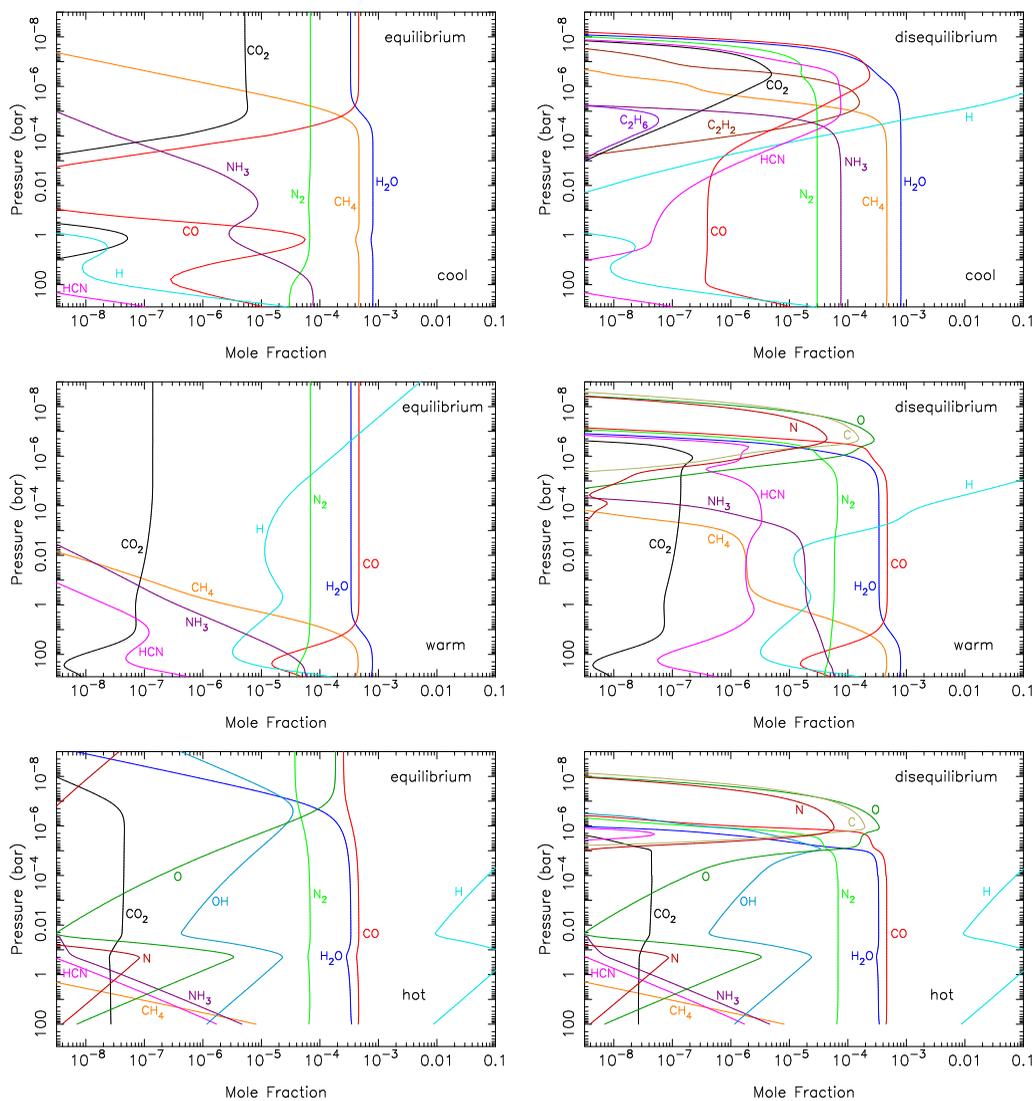

\begin{tabular}{ll}
{\includegraphics[angle=-90,clip=t,scale=0.28]{fig3_topleft.ps}} 
& 
{\includegraphics[angle=-90,clip=t,scale=0.28]{fig3_topright.ps}} 
\\
{\includegraphics[angle=-90,clip=t,scale=0.28]{fig3_middleleft.ps}} 
& 
{\includegraphics[angle=-90,clip=t,scale=0.28]{fig3_middleright.ps}} 
\\
{\includegraphics[angle=-90,clip=t,scale=0.28]{fig3_bottomleft.ps}} 
& 
{\includegraphics[angle=-90,clip=t,scale=0.28]{fig3_bottomright.ps}} 
\\
\end{tabular}
\caption{Mole fraction (volume mixing ratio) profiles for our generic cool (top), warm (middle), and 
hot (bottom) exoplanets, assuming thermochemical equilibrium (left) or thermochemical and photochemical 
kinetics and transport (right).  All models assume a solar elemental composition, and the transport 
models assume a uniform eddy diffusion coefficient of 10$^9$ $\cmtwo$ $\smone$.  Note the decrease in the 
the importance of CH$_4$ and NH$_3$ going from the ``cool'' to the ``hot'' exoplanet.  
Online version is in color.\label{figmix}}
\end{figure}

Examples of the influence of atmospheric temperatures on the photochemical composition are shown in
Figs.~\ref{figtemp} \& \ref{figmix}.  First, we present the thermal profiles of three hypothetical ``hot,'' 
``warm,'' and ``cool'' exoplanets in Fig.~\ref{figtemp}, along with some important chemical-equilibrium 
boundaries.  Although these temperatures are based on published profiles for specific exoplanets 
[63,110-111], 
we use them generically here for the photochemical models shown in 
Fig.~\ref{figmix}, which all assume a solar composition of elements and an arbitrary constant-with-altitude 
eddy diffusion coefficient of 10$^9$ $\cmtwo$ $\smone$.  The assumed incident stellar flux in the models, 
however, remains consistent with the adopted thermal structure, so that the hot planet is receiving a larger 
overall incident flux.  Note from Fig.~\ref{figtemp} that the thermal 
profile of ``cool'' exoplanet lies in a regime in which $\CHfour$ is the dominant carbon species throughout 
much of the atmosphere except at very low pressures, whereas the profile transitions from the $\NHthree$ 
regime to the $\Ntwo$ regime deeper in the atmosphere; the thermal profile of the ``warm'' exoplanet lies in 
the $\Ntwo$-dominated regime, but crosses the equal-abundance boundary between CO and $\CHfour$ near 10 
bar; and the thermal profile for the ``hot'' planet lies solidly within the $\Ntwo$ and CO stability regimes.  
The location of the thermal profile within these chemical stability boundaries controls what ``parent'' 
molecules are available to initiate photochemistry.  Figure \ref{figmix} then shows the expected composition 
from the consideration of either thermochemical equilibrium (left side of plot) or disequilibrium chemistry 
due to thermochemical and photochemical kinetics and vertical transport (right side of plot), using the 
models described in Moses et al.~\cite{moses11,moses13}.  

One obvious consequence of photochemistry on all of these close-in exoplanets, regardless of the thermal 
structure, is the huge production rate of H at high altitudes (due to H$_2$ photolysis, $\Htwo$ thermal 
decomposition in the planet's expected hot thermosphere, and H$_2$O photolysis and subsequent catalytic 
destuction of $\Htwo$). The atomic H then diffuses downward to affect the chemistry of other molecules.  
The catalytic $\HtwoO$ photolysis source of H was first described by \cite{liang03}.  Similar 
catalytic destruction mechanisms involving CH$_4$ and NH$_3$ \cite{moses11} operate in the cool and warm 
exoplanet atmospheres where these molecules are more abundant.  Atomic H is expected to become the 
dominant atmospheric constituent at relatively high ($\sim$microbar) pressures, with the exact transition 
from $\Htwo$ to H depending on the soft x-ray and EUV flux from the host star, thermospheric temperatures, 
and hydrodynamic winds \cite{garcia07}.  Hydrodynamic descriptions and ion chemistry are needed to 
accurately model the chemical behavior at altitudes above $\sim$1 $\mu$bar on highly irradiated 
planets \cite{yelle04,garcia07}, so the high-altitude results presented here are phenomenological 
only.  Neutral O, C, and N from the photodestruction of CO, N$_2$, and $\HtwoO$ are undoubtedly going 
to be important at high altitudes on highly irradiated planets, but hydrodynamic winds likely drag these 
atomic constituents to higher altitudes than are shown in this hydrostatic model.

On very hot exoplanets, like our ``hot'' generic giant planet shown in Fig.~\ref{figmix}, photochemistry 
is relatively unimportant 
[see also 62-63,65].  
Photolysis of CO and $\Ntwo$ leads 
to the high-altitude production of atomic species like C and N that are not predicted to be important 
under equilibrium conditions.  Such atomic species (as well as O) and their ions could be observable in UV 
transit observations (e.g., \cite{vidal04}).  At high altitudes, hydroxyl radicals (OH) are also produced 
from $\HtwoO$ photolysis or reaction with atomic H, and HCN is produced from CO and $\Ntwo$ photolysis
through schemes such as are described at the beginning of Section 3.5 of Moses et al.~\cite{moses11}, but 
neither of these species remain stable in regions where atomic H dominates.  Some production of NO (peaking 
at ppm abundances) and even O$_2$ (peaking at 10 ppb abundances) occurs, but these species are confined to a 
narrow altitude region where CO and $\Ntwo$ photolysis is effective (i.e., before CO and N$_2$ self shield) 
and will not be abundant enough to be observable for solar-composition atmospheres.  In fact, virtually all 
of the interesting photochemistry is confined to this narrow altitude region where the CO and N$_2$ photolysis 
rates peak.  Although $\HtwoO$ photolysis continues to much deeper levels, the photolysis products tend 
to recycle rapidly back to $\HtwoO$ in the $\Htwo$-dominated atmosphere.  Temperatures are high enough 
throughout the bulk of the infrared photosphere at $\sim$0.0001-1 bar such that kinetic reactions 
can maintain equilibrium, despite the perturbing influence of photolysis of $\HtwoO$ and other less-abundant 
species.  For the hottest planets, the reaction rates are fast enough that vertical transport-induced 
quenching is not important either, for reasonable assumptions about vertical eddy diffusion coefficients.  
If quenching occurs at all, it occurs at high altitudes (low pressures), where the abundance of 
photochemically active molecules like NH$_3$ and CH$_4$ have very low equilibrium abundances and so 
are not major players in the subsequent chemistry.  Thermochemical equilibrium is therefore a reasonable 
assumption for the atmospheric composition of the hottest hot Jupiters, although horizontal quenching may 
still occur if longitudinal thermal gradients are expected to be large (i.e., the terminator abundances 
may not necessarily be in equilibrium).  There is no specific critical temperature at which disequilibrium 
processes cease to be important, as the disequilibrium species gradually phase out with increasing 
temperature, and the results depend on transport processes as well, but both photochemistry and 
transport-induced quenching become less and less important with increasing effective temperature of 
the planet and can be safely ignored for photospheric temperatures $\gta$ 2000 K.

For our ``warm'' generic exoplanet shown in Fig.~\ref{figmix}, both photochemistry and transport-induced 
quenching are more important than on the hotter exoplanet.  Although the relatively stable molecules CO, 
H$_2$O, and N$_2$ remain the dominant carriers of O, C, and N, the quench points for CH$_4$-CO and 
N$_2$-NH$_3$ interconversion are deeper in the atmosphere (assuming ``reasonable'' transport parameters)
where the NH$_3$ and CH$_4$ mole fractions are relatively large.  Once quenched, these species are mixed 
upward into regions where they can be photochemically destroyed.  Neither NH$_3$ nor CH$_4$ is stable 
in the large background bath of H atoms diffusing down from higher altitudes; moreover, $\NHthree$
is photolyzed at longer wavelengths than the more abundant CO, $\HtwoO$, and N$_2$ such that NH$_3$ 
is not shielded against photolysis, and the longer-wavelength UV photons penetrate to deeper atmospheric 
levels where ammonia photodissociation is important.  Ammonia photolysis and subsequent catalytic destruction 
of H$_2$ \cite{moses11} adds an additional source of H atoms down to $\sim$10 mbar in the atmosphere 
of our warm exoplanet, which leads to the photosensitized destruction of CH$_4$, and subsequent coupled 
photochemistry of NH$_3$ and CH$_4$.  Some of the nitrogen from the photodestruction of NH$_3$ ends up 
in N$_2$ (with intermediates like N and even NO) through schemes such as 
\begin{eqnarray}
2\, ( \NHthree \, + \, h\nu \, & \rightarrow & \, \NHtwo \, + \, \Hatom ) \nonumber \\
\HtwoO \, + \, h\nu \, & \rightarrow & \, \OH \, + \, \Hatom \nonumber \\
2\, ( \NHtwo \, + \, \Hatom \, & \rightarrow & \, \NH \, + \, \Htwo ) \nonumber \\
2\, ( \NH \, + \, \Hatom \, & \rightarrow & \, \N \, + \, \Htwo ) \nonumber \\
\N \, + \, \OH \, & \rightarrow & \, \NO \, + \, \Hatom \nonumber \\
\N \, + \, \NO \, & \rightarrow & \, \Ntwo \, + \, \Oatom \nonumber \\
\Oatom \, + \, \Htwo \, & \rightarrow & \, \OH \, + \, \Hatom \nonumber \\
\OH \, + \, \Htwo \, & \rightarrow & \, \HtwoO \, + \, \Hatom \nonumber \\
2\, \Hatom \, + \, \M \, & \rightarrow & \, \Htwo \, + \, \M \nonumber \\
\noalign{\vglue -10pt}
\multispan3\hrulefill \nonumber \cr
\Net \ \ 2\, \NHthree \, & \rightarrow & \, \Ntwo \, + \, 3\,\Htwo  . \\
\end{eqnarray}
Some of the nitrogen ends up in HCN through schemes such as 
\begin{eqnarray}
\NHthree \, + \, h\nu \, & \rightarrow & \, \NHtwo \, + \, \Hatom  \nonumber \\
\NHtwo \, + \, \Hatom \, & \rightarrow & \, \NH \, + \, \Htwo  \nonumber \\
\NH \, + \, \Hatom \, & \rightarrow & \, \N \, + \, \Htwo \nonumber \\
\Hatom \, + \, \CHfour \, & \rightarrow & \, \CHthree \, + \, \Htwo \nonumber \\
\N \, + \, \CHthree \, & \rightarrow & \, \HtwoCN \, + \, \Hatom \nonumber \\
\HtwoCN \, & \rightarrow & \, \HCN \, + \, \Hatom \nonumber \\
\noalign{\vglue -10pt}
\multispan3\hrulefill \nonumber \cr
\Net \ \ \NHthree \, + \, \CHfour \, & \rightarrow & \, \HCN \, + \, 3\,\Htwo  . \\
\end{eqnarray}
In fact, HCN is the ultimate product of the coupled NH$_3$-CH$_4$ photochemistry, and it becomes 
an important atmospheric constituent in the infrared photosphere of all but the hottest 
hot Jupiters.  As discussed in the previous section, HCN is also an important thermochemical 
product that can form in ``warm'' exoplanets from the kinetics of quenched CH$_4$ and NH$_3$ 
\cite{moses11} via a psuedo-equilibrium that continues between these species despite the 
cessation of reactions involved with the N$_2$-NH$_3$ and CO-CH$_4$ quenching.

Hydrogen cyanide is therefore an important disequilibrium product that has been largely ignored 
to date by exoplanet spectral modelers (with the exception of 
[62-64,113-114] 
despite the fact that HCN will be a 
major constituent whenever atmospheric temperatures are cool enough that CH$_4$ and NH$_3$ quench at 
significant abundances.  For atmospheres with near-solar elemental compositions, HCN is likely 
to be even more abundant than molecules like CO$_2$ that are typically considered in spectral models
(see Fig.~\ref{figmix}).

Acetylene (C$_2$H$_2$) is another potentially important disequilibrium product on extrasolar giant 
planets 
[55,60-66].  
It is not formed efficiently in the ``warm'' exoplanet model presented here, due to the relatively high 
stratospheric temperatures adopted for this planet, but it can form in abundance on cooler planets 
through both CO photolysis at high altitudes, and through schemes initiated by the reaction of 
photochemically produced H with CH$_4$ \cite{moses11}.  Acetylene tends to be more prevalent at 
high altitudes on these planets because at higher pressures it is converted efficiently back 
to methane (and/or to the more hydrogen-saturated ethane on cooler giant planets).

The photochemistry of oxygen compounds on our generic ``warm'' exoplanet is comparatively less 
interesting than that of carbon and nitrogen.  Photolysis of carbon monoxide occurs only at very 
high altitudes, and CO with its strong carbon-oxygen bond, is kinetically stable at lower altitudes 
once the EUV photons responsible for photodissociation have all been absorbed higher up.  Water 
photolysis continues to lower altitudes, but the photolysis products are efficiently converted back 
to H$_2$O as long as sufficient H$_2$ is present.  Thus, both CO and H$_2$O follow their equilibrium 
profiles throughout the 0.0001-1 bar photospheric region.  Carbon monoxide maintains a kinetic 
equilibrium with CO and H$_2$O via the reaction CO$_2$ + H $\rightleftharpoons$ CO + OH throughout 
most of the photosphere.  Some minor excess CO$_2$ is produced photochemically at high altitudes 
due to CO and water photolysis, but the large background H abundance allows efficient recycling back 
to CO.  At lower altitudes, CO$_2$ can be photolyzed by longer-wavelength UV photons to form primarily 
CO + O($^1$D) or CO + O, but the O($^1$D) and O react with H$_2$ to form OH, and the OH reacts with 
the CO to reform the CO$_2$, so CO$_2$ is photochemically stable, again as long as sufficient H$_2$ 
is present.  Carbon dioxide therefore maintains its equilibrium abundance throughout the 0.0001-1 bar 
photospheric region in warm giant-planet atmospheres.

Photochemistry within the atmosphere of our ``cool'' exoplanet shown in Fig.~\ref{figmix} is 
fundamentally different from that of warmer planets whose photosphere resides within the CO stability 
field (e.g., \cite{line11gj436b,miller-ricci12}).  Methane is now the stable parent molecule for 
the subsequent carbon photochemistry, and the resulting net production rates for complex hydrocarbons 
are larger than for the hotter planets.  As is discussed by Line et al.~\cite{line11gj436b}, 
C$_2$H$_x$ species become important photolysis products at high altitudes, although they tend to be 
converted back to methane at lower altitudes.  Coupled NH$_3$-CH$_4$ photochemistry as described above 
and in Moses et al.~\cite{moses11} and Line et al.~\cite{line11gj436b} is effective, and HCN becomes a major 
photochemical product both locally at various altitudes and from a column-integrated standpoint.  
Consistent with Line et al.~\cite{line11gj436b}, we expect the CH$_4$ to remain stable throughout the 
0.0001-1 bar infrared photosphere on cooler giant planets.  Coupled $\HtwoO$-$\CHfour$ photochemistry 
converts the methane to CO at higher altitudes, but these processes are less effective at lower altitudes 
due to a lack of production of reactive radicals once H$_2$O (and CH$_4$) photolysis shuts down due to 
$\HtwoO$ self-shielding at $\sim$10$^{-4}$ bar and once H has been scavenged back into stable 
hydrogen-saturated molecules.  There are numerous effective pathways for this coupled $\HtwoO$-$\CHfour$
photochemistry: (a) atomic O produced from $\HtwoO$ photolysis can react with $\CHthree$, $\CtwoHtwo$, 
$\CthreeHtwo$, and $\CthreeHthree$ to form carbon-oxygen bonded species that eventually form CO, 
(b) CH can react with water to form $\HtwoCO$ and ultimately CO, and (c) C reacts with NO produced 
from coupled N$_2$-H$_2$O photochemistry to form N + CO.  The carbon-bearing radicals and molecules 
involved with the above schemes derive ultimately from the reaction of methane with H released from 
water photolysis.  Some CO$_2$ is produced from this coupled CH$_4$-H$_2$O photochemistry, but CO$_2$ 
itself is photolyzed to produce CO, and the reaction CO + OH $\rightarrow$ CO$_2$ + H cannot recycle 
the CO$_2$ fast enough at these temperatures to maintain the CO-H$_2$O-CO$_2$ equilibrium.  Carbon 
dioxide is therefore not a major photochemical product in this cool-exoplanet model.

Transport-induced quenching does enhance the abundance of CO in the IR photosphere of our ``cool'' 
exoplanet, and photochemical production of CO occurs at high altitudes, but the column abundance 
of CO never rivals that of methane in the infrared photosphere.  We therefore agree with the 
conclusions of Line et al.~\cite{line11gj436b} that disequilibrium processes cannot remove 
methane in favor of CO and/or CO$_2$ on cooler, solar-composition, giant planets like GJ 436b.

Due to the prevalence of both NH$_3$ and CH$_4$ in cooler giant exoplanets, photochemical production 
of complex hydrocarbons and nitriles is favored in the atmospheres of these cool planets as compared 
with hotter planets.  Benzene (C$_6$H$_6$) and cyanoacetylene (HC$_3$N), for example (not shown in 
Fig.~\ref{figmix} for reasons of clarity), both achieve peak mole fractions of $\sim$2$\scinot-7.$ 
in our generic cool exoplanet.  While these species will not condense at atmospheric conditions 
relevant to this planet, continued production of refractory species not considered in this model 
is likely, and the formation of high-altitude photochemical hazes is therefore more probable on 
these cooler exoplanets 
[cf. 55,62,115]. 

\section{Sensitivity of disequilibrium chemistry to transport parameters and bulk elemental ratios}

We discussed the sensitivity of disequilibrium chemistry and composition on hot Jupiters to atmospheric 
temperatures in the previous section; the sensitivity to vertical transport parameters 
like eddy diffusion coefficients ($K_{zz}$) is discussed in detail in 
[62,64,115]. 
The composition is mainly sensitive to the $K_{zz}$ profile through its influence on the quench pressure 
in transport-induced quenching, and hence through its effect on the abundance of quenched species.  If the 
quenched species are photochemically active at higher altitudes, there can be important consequences 
with respect to the resulting abundance of photochemical products like HCN, C$_2$H$_x$, and complex 
hydrocarbons and nitriles.  The greater the strength of atmospheric mixing at the quench pressures 
(i.e., the larger the $K_{zz}$ values), the deeper the quench points will be, and the greater the 
resulting abundance of quenched disequilibrium species like CH$_4$ and NH$_3$ on warm Jupiters.  
That also leads to greater lower-atmospheric abundances of kinetically dependent species like HCN and 
C$_2$H$_2$.  Similarly, on cooler exoplanets where CH$_4$ is the dominant carbon species, a greater 
$K_{zz}$ at the CO-$\CHfour$ quench point leads to a greater quenched abundance of CO.  Quenching of 
NH$_3$ and/or N$_2$ is also very sensitive to $K_{zz}$ values on cooler exoplanets, where smaller $K_{zz}$ 
values could push the quench point into the $\Ntwo$-dominated regime, allowing N$_2$ rather than NH$_3$ 
to be the dominant nitrogen component in the upper atmosphere, with a corresponding reduction in the 
column abundances of all disequilibrium nitrogen species.

The strength of vertical mixing in the upper atmosphere also affects the abundance of photochemically 
produced constituents.  The greater the $K_{zz}$ values in the upper atmosphere, the higher the altitude 
to which the parent molecules can be carried, with a corresponding increase in the column abundance 
of species that are produced photochemically at high altitudes.  The gradient of the $K_{zz}$ profile 
can affect how rapidly photochemically produced species are transported downward to their 
thermochemical destruction regions and thus the overall column 
abundance below the high-altitude production region \cite{moses05,moses00}.

Bulk elemental abudances within the exoplanet atmosphere also affect the predicted equilibrium and 
disequilibrium compositions.  The metallicity of the atmosphere, for example, can strongly influence the 
expected chemical-equilibrium abundances of heavy molecules 
[42,60-62,64,115,117,124].  
All molecules that contain heavy elements tend to exhibit an increase in abundance when the metallicity is 
increased.  However, the increase in metallicity (all other parameters being equal) leads to a greater 
increase in CO in comparison to CH$_4$ and a greater increase in $\Ntwo$ in comparison with NH$_3$, and 
molecules with three or more heavy nuclei, like CO$_2$, are even more sensitive to metallicity.  The 
above references all emphasize the potential importance of CO$_2$ as a probe of the atmospheric metallicity 
on exoplanets, as an increase in metallicity by a factor of $x$ tends to increase the abundance of 
CO$_2$ by a factor of $x^2$.  Photochemical products such as NO, C$_2$H$_x$, or HCN that depend on the 
abundance of the augmented heavy species tend to also become more abundant as the metallicity increases, 
although the response of disequilibrium species can sometimes be subtler.  For example, Moses et al.~\cite{moses11} 
find that the quenched $\CHfour$ mole fraction in the infrared photosphere of their HD 189733b model 
actually decreases when the metallicity is increased by a factor of 10 because the $\CHfour$ $\rightarrow$ 
CO conversion schemes become more effective when the H$_2$O and CO abundances increase, such that 
$\CHfour$ quenches at higher altitudes where the mole fraction is smaller.  

The bulk atmospheric elemental ratios can also strongly affect the composition, both from equilibrium- 
and disequilibrium-chemistry standpoints.  The effect of the C/O ratio on the equilibrium composition 
has been discussed by
[63,65,111,114,118-120];  
the disequilibrium chemistry 
consequences have been discussed by \cite{moses13,koppa12}.
The C/O ratio strongly influences the abundance of spectrally active molecules like CO$_2$, H$_2$O, CH$_4$, 
HCN, and $\CtwoHtwo$, particularly on hot Jupiters that are warm enough that CO is expected to be the 
dominant form of carbon.  For C/O ratios $<$ 1 in thermochemical equilibrium at photospheric pressures, 
methane rapidly decreases in importance with increasing temperature as more and more of the carbon is 
sequestered in CO, and species like $\CtwoHtwo$ and HCN are unimportant at all temperatures.  At C/O 
ratios $>$ 1, methane is less temperature sensitive, and species like HCN and $\CtwoHtwo$ rapidly gain 
in importance with increasing temperature such that they eventually become the dominant carriers of 
carbon behind CO.  The resulting effects on the atmospheric spectrum can be major
[63,111,114,118]. 
Other elemental ratios like N/O and C/N will have 
similar interesting effects that have been less well studied.

\begin{figure}
\includegraphics[clip=t,scale=0.7]{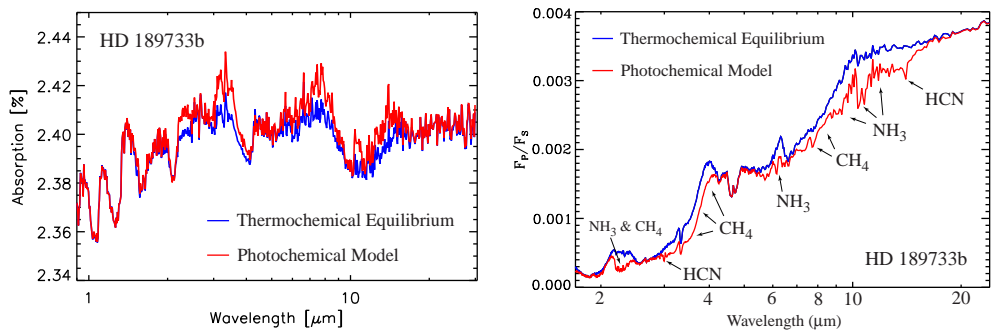}
\caption{(Left) Synthetic transit spectra for solar-composition HD 189733b models assuming thermochemical 
equilibrium (blue) or considering thermochemical/photochemical kinetics and transport (red).  Absorption 
depth is plotted as the square of the apparent planet-to-star radius ratio (figure is modified from 
\cite{moses11}, with spectral calculations from J. J. Fortney). (Right) Synthetic eclipse emission 
spectra for solar-composition HD 189733b models assuming thermochemical equilibrium (blue) or considering 
thermochemical/photochemical kinetics and transport (red), plotted in terms of the flux of the planet 
divided by the flux of the star (figure is modified from \cite{moses11}, with spectral calculations from 
C. A. Griffith).  Disequilibrium molecules responsible for absorption bands in the emission spectrum 
are labeled specifically.  Online version is in color.\label{figspecphot}}
\end{figure}

\section{Observational consequences of disequilibrium chemistry}

Both transport-induced quenching and photochemistry will affect the spectral properties of exoplanet 
atmospheres through the influence on the composition.  Some of these spectral consequences for extrasolar 
giant planets have been discussed by 
[17,62-64,67,69,72,86,113-114,122-124,126].  
On the whole, the effects can be minor (see Fig.~\ref{figspecphot}) because the disequilibrium processes 
tend not to affect the dominant heavy constituents within the 0.0001-1 bar region (e.g., where CO and 
$\HtwoO$ dominate on warmer hot Jupiters, and where $\HtwoO$ and $\CHfour$ dominate on cooler hot Jupiters), 
and these dominant species --- water in particular --- control the spectral properties throughout most of the 
infrared.  The disequilibrium effects are then expected to be mostly apparent within infrared windows 
where water does not strongly absorb, such as the $\sim$1.4-1.8, 2.0-2.5, 3-5, and 
8-16 $\mu$m regions.

One effect that can be less subtle is the consequence of CO-$\CHfour$ quenching on hotter giant planets
with strong longitudinal thermal gradients, such that CO is expected to be the dominant 
thermochemical-equilibrium carbon phase on the dayside, but CH$_4$ is expected to be the dominant carbon 
phase at the terminators and/or nightside.  As was emphasized by Cooper \& Showman \cite{coop06}, 
CO-$\CHfour$ quenching (both vertical and horizontal) can prevent CH$_4$ from forming in significant 
quantities in the cooler regions where equilibrium arguments would expect it to be significant, and the 
consequences for transit spectra and for predicted spectral variations as a function of orbital phase 
can be major 
[67,69,72,124-125].  
If CO-$\CHfour$ quenching in high-temperature regions prevents CH$_4$ from being present in significant 
quantities in cooler atmospheric regions, the resulting planetary spectrum can exhibit excess opacity in 
the $\sim$2.3-2.4 and 4.5-4.9 $\mu$m wavelength regions, where CO has strong vibrational bands, and 
reduced opacity in the $\sim$2.15-2.45, 3.15-3.45, and 7.2-8.2 $\mu$m wavelength regions, 
where $\CHfour$ has strong vibrational bands.  Transit and nightside spectra will be particularly 
sensitive to such CO quenching effects.  

On the other hand, vertical quenching of CH$_4$ can potentially supply methane to both the dayside and 
terminator photospheres on ``warm'' Jupiters in excess of what would be expected in equilibrium, 
particularly if the CO-$\CHfour$ quench point is at or below the base of the ``photosphere.'' If 
horizontal thermal gradients still exist at the CO-$\CHfour$ quench point, the final photospheric 
$\CHfour$ abundance on planets like HD 189733b and HD 209458b will be a complicated function of 
both vertical and horizontal quenching; however, in essence, vertical quenching will supply methane to 
the photosphere, and strong horizontal winds and horizontal quenching will help homogenize that methane
as a function of longitude.  Starting from realistic thermal profiles from general circulation models 
\cite{showman09}, Moses et al.~\cite{moses11} find that even in the absence of horizontal quenching, 
vertical transport-induced quenching supplies a quenched CH$_4$ mole fraction to the photosphere that 
only differs by a factor of a few between the terminator and dayside models, compared with the large 
difference of those disequilibrium predictions in comparison with equilibrium expectations (see
Fig.~\ref{quenchcomp}).  Therefore, transport-induced quenching is expected to provide excess 
photospheric methane in comparison with equilibrium predictions for both HD 189733b and HD 209458b, 
with some notable spectral consequences, particularly for the cooler HD 189733b (see 
\cite{moses11} and Fig.~\ref{figspecphot}).  The main consequence is an increase in absorption at 
$\sim$2.15-2.45, 3-4, and 7-9 $\mu$m from what would be expected based on equilibrium compositions.

Quenching of $\NHthree$ on hot Jupiters over a wide range of temperatures can influence the spectrum 
by providing additional opacity in the $\sim$1.5, 2.0, 2.2-2.3, 3.0, 5.5-7.0, and 8-12 $\mu$m regions, 
where NH$_3$ has strong bands.  Ammonia and methane quenching 
provide a thermochemical-kinetics source of HCN, which photochemistry can also enhance in the upper 
atmosphere.  Disequilibrium HCN then provides excess absorption in the $\sim$1.53, 1.85, 2.8-3.1, 
6.5-7.5, and 13-16 $\mu$m wavelength regions, with particularly strong contributions at $\sim$3 and 
14 $\mu$m.  As can be seen from Fig.~\ref{figspecphot}, these disequilibrium nitrogen species can 
have a notable influence on the exoplanetary spectra within the water-absorption windows, and we 
encourage the inclusion of HCN and NH$_3$ in exoplanet spectral models --- and encourage the 
acquisition of high-temperature line parameters for NH$_3$ and HCN so that spectral modelers can 
reliably include these species in their calculations.

Acetylene is another photochemical product that can have a potentially important influence on the 
spectra of cooler planets.  Given its predicted abundance on HD 189733b \cite{line10,moses11,venot12}, 
$\CtwoHtwo$ will not have much of an influence on spectra from that planet (see \cite{moses11} and 
Fig.~\ref{figspecphot}), being overshadowed by the more abundant HCN and CH$_4$ in the $\sim$3 and 7-8 $\mu$m 
regions where $\CtwoHtwo$ has bands, but acetylene also has several distinct bands in the 1-3 $\mu$m region
and the 12.5-15 $\mu$ region (with a strong $\nu_5$ Q branch at $\sim$13.7 $\mu$m) that might 
provide a unique spectral signature for cooler planets where $\CtwoHtwo$ is expected to be more abundant.  
Acetylene will also be a more important constituent for atmospheres with higher C/O ratios
[63,65,114],
and if there are reasons to expect a high C/O ratio for any given 
planet 
[e.g., 63,111,114], 
both HCN and $\CtwoHtwo$ should be considered 
in spectral calculations.

\section{Current evidence for disequilibrium chemistry on exoplanets}

In the previous section, we discussed some of the expected observational consequences of disequilibrium 
chemistry on the spectral properties of exoplanets, but it remains to be demonstrated whether actual 
observations provide any evidence for these effects.  Here, we look at some of the existing claims 
for disequilibrium compositions in the infrared photospheres of extrasolar giant planets and discuss 
whether photochemistry and transport-induced quenching can account for the observed behavior or 
whether other processes or atmospheric characteristics must be responsible for that behavior.  

Disequilibrium chemistry in the troposphere/stratosphere has been suggested to bring about 
the following observed hot Jupiter or hot Neptune characteristics:

\begin{itemize}
\item {\em Weak CH$_4$ absorption on young, directly-imaged giant planets.}  Near-infrared spectra and 
narrow-band photometry suggest that $\CHfour$-CO quenching occurs on several of the better-studied 
directly imaged exoplanets like the those in the HR 8799 system 
[13-15,17,20-21] 
and 2M1207b \cite{barman112m}.  
Quenching of CO at much greater than equilibrium abundances is expected to have important spectral 
consequences on such planets \cite{fort08young}, and indeed is very likely from theoretical grounds, 
leading to greatly enhanced CO abundances and possible CO/CH$_4$ ratios greater than unity 
\cite{barman11hr8799}.  Although clouds and potential non-solar metallicities can complicate the 
interpretation \cite{currie11,barman11hr8799,madhu11hr8799,marley12}, the lack of evidence for 
strong methane absorption does seem to indicate the occurrence of disequilibrium transport-induced 
quenching of CO and CH$_4$ on these planets.  This conclusion is further reinforced by recent 
high-spectral-resolution observations of HR 8799c \cite{konopacky13} that show features due to 
water and CO but not methane.

\item {\em Near-IR detections of CH$_4$ on HD 189733b and HD 209458b.}  Swain et al.~\cite{swain08hd189} 
report the detection of CH$_4$ on HD 189733b from transit spectroscopy obtained from the NICMOS 
instrument on the {\em Hubble Space Telescope\/} (HST).  The relatively large amount of methane 
inferred from these observations \cite{swain08hd189,madhu09} is in excellent agreement with predictions 
that include disequilibrium transport-induced quenching of $\CHfour$ \cite{line10,moses11,visscher11}.  
However, the lack of detection of methane (and the corresponding low upper limit) on the dayside of 
the planet with the same instrument \cite{swain09hd189} is in conflict with disequilibrium chemistry 
predictions that include vertical transport \cite{line10,moses11,visscher11,venot12}, making the case 
for methane via disequilibrium processes inconclusive on HD 189733b.  Although photochemical and 
thermochemical kinetics processes are expected to remove CH$_4$ from the upper portions of the dayside 
atmosphere on HD 189733b, the overall column abundance should not be affected much, and transport-induced 
quenching (both vertical and horizontal) should ensure dayside and terminator methane abundances 
that differ by only a factor of a few.  By the same token, the large dayside abundance of methane 
inferred from HST/NICMOS eclipse observations of HD 209458b\cite{swain09hd209} seems too large to be 
explained by 
transport-induced quenching \cite{moses11}; if the spectral signatures are robust, we suggest that 
alternative (yet to be identified) molecules should be considered as possible candidates for the 
absorption, or that the atmosphere has a C/O ratio greater than solar.  With regard to the latter 
point, we note that although the {\em Spitzer\/} infrared photometric data from eclipse cannot 
provide meaningful constraints on the CH$_4$ abundance of HD 209458b \cite{madhu09}, the relatively 
low water abundance inferred from both the HST/NICMOS spectra and {\it Spitzer\/} photometric data 
\cite{swain09hd209,madhu09} is also consistent with an inferred high C/O ratio on HD 209458b, but 
the relatively high derived CO$_2$ abundance from HST/NICMOS data \cite{swain09hd209} is not.  We 
conclude that the methane detections on both HD 189733b and HD 209458b do not provide any 
unambiguous evidence for disequilibrium $\CHfour$ quenching on these planets.  Further near-IR 
observations, especially from space-based platforms like the {\em James Webb Space Telescope\/} (JWST)
\cite{gardner06,clampin11}, FINESSE \cite{swain12}, or EChO \cite{tinetti12} could help resolve 
this issue.

\item {\em Orbital phase curves for HD 189733b at 3.6 and 4.5 $\mu$m.}  Knutson et al.~\cite{knutson12} report 
phase-variation observations for HD 189733b over a full orbit that may have implications with respect to 
CO-$\CHfour$ transport-induced quenching.  They find that 3D circulation models that assume chemical 
equilibrium have spectral signatures that compare well to {\em Spitzer\/} photometric channel 
fluxes for conditions in the dayside atmosphere at or near the secondary eclipse, but that the 
models considerably overpredict the flux in the 4.5 $\mu$m channel on the nightside.  Knutson et
al.~\cite{knutson12} suggest that this nightside behavior at 4.5 $\mu$m, plus generally insufficient 
model absorption at 4.5 $\mu$m in transit simulations, could be a signature of CO quenching, such 
that more CO exists in the terminator and nighttime atmosphere than is predicted from equilibrium 
models.  This trend is indeed the expected one for CO-CH$_4$ quenching, but it is not clear 
from the information provided in \cite{knutson12} how much of the carbon is tied up in CH$_4$ in their 
nightside equilibrium model atmosphere and/or whether the magnitude of the quenching effect would be 
sufficient to produce the observed behavior (and note that disequilibrium models of Line et 
al.~\cite{line10}, Moses et al.~\cite{moses11,moses13}, and Venot et al.~\cite{venot12} have a lower CH$_4$/CO 
ratio in general than is considered in the Knutson et al.~\cite{knutson12} quench discussion), nor is 
it clear whether other bulk atmospheric model parameters such as a non-solar metallicity or C/O ratio 
could produce the described behavior.  Therefore, although the observations are certainly suggestive 
of disequilibrium quenching effects, additional models that explore a more complete range of parameter 
space, as well as higher-spectral-resolution observations in the 1-5 $\mu$m region that could help separate 
contributions from CO and CH$_4$, would help place the claim of disequilibrium effects on a firmer 
foundation.  The observed phase-curve behavior for HD 189733b at 3.6 $\mu$m is also interesting in that the 
minimum-to-maximum brightness temperature range over the orbit is greater at 3.6 $\mu$m than at 4.5 
$\mu$m \cite{knutson12}, despite equilibrium expectations that the 3.6 $\mu$m band should have weaker 
average opacity than the 4.5 $\mu$m band and thus probe deeper atmospheric levels where the thermal 
structure is not as variable.  Additional disequilibrium opacity sources such as HCN or quenched 
methane could be causing the 3.6 $\mu$m channel to probe higher altitudes; however, the fact that 
the eclipse depth at 3.6 $\mu$m corresponds to a larger brightness temperature than at 4.5 $\mu$m 
\cite{knutson12} suggests that the 3.6 $\mu$m channel does indeed probe deeper levels on the dayside, 
as expected, and the larger phase variations at 3.6 $\mu$m then might involve opacity changes at deep 
levels, which are not predicted from disequilibrium models.  Again, further explorations of parameter 
space are required before the interesting observed behavior can be better understood.

\item {\em Large inferred CO$_2$ abundance on HD 189733b.}  The HST/NICMOS secondary-eclipse observations 
of Swain et al.~\cite{swain09hd189} seem to require an unexpectedly large abundance of CO$_2$ in 
the dayside atmosphere of HD 189733b \cite{madhu09,lee12,line12opti}, leading to the suggestion that 
the CO$_2$ is supplied by photochemical processes \cite{madhu09}.  However, thermo/photochemical 
kinetics and transport models \cite{line10,moses11,venot12} predict only small high-altitude increases 
due to photochemistry in the otherwise low expected CO$_2$ equilibrium abundance.  In fact, retrievals 
that are based on the Swain et al.~\cite{swain09hd189} data \cite{madhu09,lee12,line12opti} suggest 
that the inferred CO$_2$ abundance is greater than that of water on HD 189733b --- a situation that 
is very unlikely in a hydrogen-dominated atmosphere even under disequilibrium conditions.  
Therefore, as is emphasized by Moses et al.~\cite{moses13}, the HST/NICMOS secondary-eclipse data, 
if robust, suggest that either (a) some other molecule is responsible for the absorption attributed to 
CO$_2$, (b) the atmospheric metallicity of HD 189733b is extremely high (e.g., several thousand times 
solar to allow CO$_2$/H$_2$O ratios greater than 1, which seems inconsistent with the high H \& He 
content indicated by the planet's mass-radius relationship \cite{marley07}), or (c) all the current 
photochemical models are missing a major mechanism that irreversibly converts H$_2$O and CO into 
CO$_2$ (which seems unlikely in a hydrogen-dominated atmosphere).  The Swain et al.~\cite{swain09hd189} 
HST/NICMOS data therefore do not provide convincing evidence for disequilibrium process on HD 189733b, 
and some other factor must be at play here.

\begin{figure}
\centering
\includegraphics[scale=0.7]{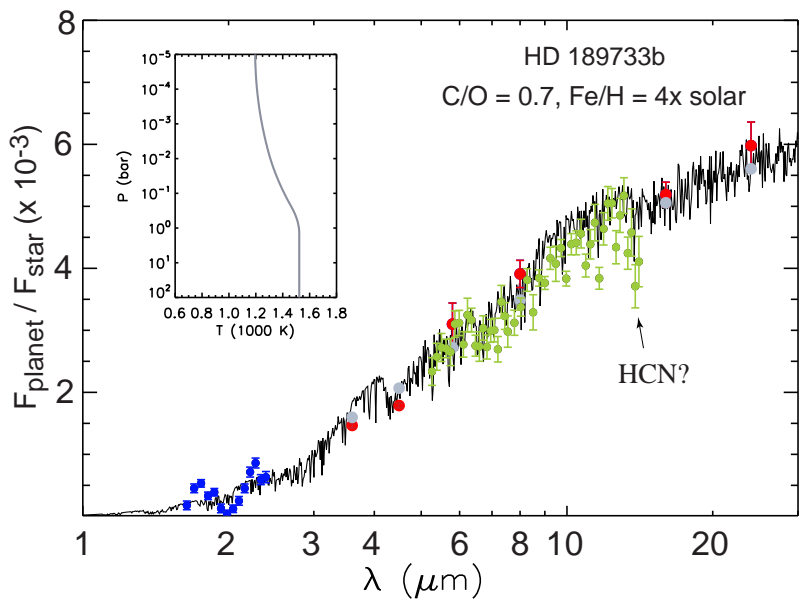}
\caption{Synthetic eclipse spectra for HD 189733b from a Moses et al.~\cite{moses13} disequilibrium 
chemistry model that assumes a C/O ratio of 0.7 and a metallicity of 4$\times$ solar (solid black 
line), compared with {\em Spitzer\/} broadband photometric points at 3.6 and 4.5 $\mu$m  
\cite{knutson12} and 5.6, 8, 16, and 24 $\mu$m \cite{knut07,knut09,charb08} (large red circles with
error bars), with {\em Spitzer\/} IRS spectra \cite{grill08} (medium green circles with error bars), and 
with HST/NICMOS spectra \cite{swain09hd189} (small blue circles with error bars).  The gray circles 
without error bars represent the model results convolved over the {\it Spitzer\/} broadband channels.  
The insert in the upper left shows the thermal profile adopted in the modeling (figure is modified 
from \cite{moses13}, with spectral calculations from N. Madhusudhan).  
Online version is in color.\label{fig14microns}}
\end{figure}

\item {\em Spitzer/IRS absorption at $\sim$14 $\mu$m on HD 189733b.}  {\em Spitzer}/IRS secondary eclipse 
spectra of HD 189733b \cite{grill08} exhibit broad features consistent with water absorption in the 
5-10 $\mu$m region on HD 189733b.  Although not discussed by Grillmair et al.~\cite{grill08}, the 
IRS spectra also exhibit a noticeable downturn at longer wavelengths that is suggestive of a 
$\sim$13.5-14.5 $\mu$m absorption band, whereas {\em Spitzer\/} IRS and MIPS broadband photometric 
observations at 16 and 24 $\mu$m \cite{charb08} jump back to higher brightness-temperature values.  
If the downturn in the flux at the longer wavelengths in the IRS spectrum 
is a real property of the atmosphere and not an observational artifact, we note from 
Fig.~\ref{fig14microns} that this behavior is reminiscent of a strong predicted $\sim$14 $\mu$m absorption 
feature due to HCN produced from thermochemical and photochemical kinetics in the disequilibrium 
models of Moses et al.~\cite{moses13} (see also \cite{madhu12coratio}).  This identification is 
certainly not definitive, and a more 
thorough investigation of model parameter space may suggest other alternatives, but we note that 
the HCN produced from disequilibrium process on HD 189733b is expected to survive throughout the 
atmosphere on both the dayside and nightside.  Searches for additional evidence for HCN on HD 189733b 
at other relevant wavelengths in transit and eclipse data might be worthwhile.

\item {\em Relative 3.6-to-4.5-$\mu$m flux ratio on GJ 436b.}  Broadband {\em Spitzer} secondary eclipse 
observations of the hot Neptune GJ 436b \cite{stevenson10}, and in particular the high observed 
flux at 3.6 $\mu$m in combination with the low flux (i.e., non-detection) at 4.5 $\mu$m, suggest 
that CO and not CH$_4$ is the dominant carbon constituent on the dayside of this cooler exoplanet, in 
serious conflict with chemical-equilibrium predictions \cite{stevenson10,madhu11gj436b}.  Stevenson et 
al.~\cite{stevenson10} and Madhusudhan \& Seager \cite{madhu11gj436b} suggest that disequilibrium 
processes are responsible for this observed behavior, with transport-induced quenching in combination 
with a high metallicity (10$\times$ solar) producing the large required atmospheric CO and $\COtwo$ 
abundances in the photosphere, and photochemistry removing the large expected large CH$_4$ abundance 
\cite{madhu11gj436b}.  However, Line et al.~\cite{line11gj436b} convincingly demonstrate that 
photochemistry cannot remove CH$_4$ from the bulk of the photosphere, and the thermal profiles 
derived from the general circulation models of Lewis et al.~\cite{lewis10} for metallicities up to 
50$\times$ solar do not predict a high-enough CO abundance at depth or anywhere else in the atmosphere 
for transport-induced quenching to supply the necessary CO mole fraction to the infrared photosphere 
\cite{richardson13}.  Therefore, if the observations are robust, some process or atmospheric property other 
than transport-induced quenching and photochemistry must be responsible for the behavior, such as an 
extremely high metallicity \cite{richardson13} and/or high-altitude clouds, unless existing photochemical 
models are missing key mechanisms that efficiently convert water and methane to CO at relevant GJ 436b 
conditions.  
\end{itemize}

The above discussions emphasize that many of the observational oddities that appear inconsistent
with equilibrium chemistry in a near-solar-composition atmosphere also remain inconsistent with 
disequilibrium chemistry in such atmospheres.  Photochemistry and transport-induced quenching are 
undoubtedly occurring on extrasolar giant planets, but the observational evidence gathered to date 
is not yet compelling, with the exception of CO quenching on young, directly imaged exoplanets.  
Certain atmospheric characteristics of hot Jupiters as described above, however, suggest that 
kinetic effects matter.  Future observations are needed to help firm up these tantalizing hints of 
disequilibrium behavior; in the process, we will gain a better understanding of the underlying 
processes that control the planet's current atmospheric composition and its possible past and 
future evolution.

\section{Conclusions}

The ability to detect and characterize atmospheres of extrasolar planets represents a phenomenal 
success story in modern astronomy.  We are, however, still in a learning phase, both in terms of the 
cutting-edge observational and analysis techniques needed to retrieve molecular abundances on 
hot Jupiters and in terms of theoretical models needed to interpret the observations.  Disequilibrium 
chemistry models play an important role in the process.  Photochemistry and transport-induced 
quenching will drive the exoplanet atmospheric compositions away from chemical equilibrium, and 
these effects will have observational consequences.

The main lessons learned to date from disequilibrium chemistry models are that transport-induced 
quenching is expected to affect the relative abundances of the carbon-bearing molecules CO and CH$_4$ 
and the nitrogen-bearing molecules N$_2$ and NH$_3$, which can have a notable impact on the transit, 
eclipse, and orbital-phase-variation observations in the spectral regions where these molecules have 
absorption bands.  If the atmospheric thermal profile crosses the stability regimes where CO-$\CHfour$ 
or N$_2$-NH$_3$ are stable in chemical equilibrium, the effects can be relatively major.  If the 
thermal profile resides solidly within one regime or another, the effects can be subtle.  Very hot 
planets will tend to kinetically maintain equilibrium, and disequilibrium effects will be confined to 
very high altitudes, except perhaps where the hotter parcels of gas are rapidly transported to cooler 
terminator or nightside regions.  On warm to moderately hot Jupiters, H$_2$O and CO will remain at 
near-equilibrium abundances throughout the infrared photosphere, but transport-induced quenching will 
increase the abundance of CH$_4$ and NH$_3$ above equilibrium expectations.  Further thermochemical and 
photochemical processing of the quenched CH$_4$ and NH$_3$ can lead to significant production of 
HCN (and in some cases C$_2$H$_2$), which can add opacity sources that fill in windows between water 
absorption bands.  Carbon dioxide is relatively unaffected by disequilibrium chemistry but remains 
a very minor atmospheric constituent on hot Jupiters unless the atmospheric metallicity is significantly
higher than solar.  On cool exoplanets where methane is expected to be stable, transport-induced
quenching can increase the expected abundance of CO, but photochemistry is not expected to remove 
methane from the troposphere and stratosphere.  Methane and ammonia photochemistry on such planets 
will result in the production of complex hydrocarbons and nitriles that might produce high-altitude 
photochemical hazes.  On exoplanets of a wide variety of temperatures, HCN and NH$_3$ will be 
important disequilibrium constituents that should not be ignored in observational analyses.

Higher-resolution infrared spectra from existing and future ground-based and space-based telescopes 
promise to provide the ``smoking guns'' needed to identify disequilibrium chemical constituents and 
their underlying kinetic controlling mechanisms.



\ack{
This work was supported by the NASA Planetary Atmospheres Program grant number 
NNX11AD64G.}

\clearpage

\end{document}